# SECONDARY ELECTRON YIELD MEASUREMENTS OF FERMILAB'S MAIN INJECTOR VACUUM VESSEL

D. J. Scott, D. Capista, K. L. Duel, R. M. Zwaska, Fermilab, Batavia, IL, U.S.A.,

S. Greenwald, W. Hartung, Y. Li, T. P. Moore, M. A. Palmer, CLASSE, Cornell University, Ithaca, NY, U.S.A.,

R. Kirby, M. Pivi, L. Wang, SLAC, Menlo Park, CA, U.S.A.

*Abstract*

We discuss the progress made on a new installation in Fermilab's Main Injector that will help investigate the electron cloud phenomenon by making direct measurements of the secondary electron yield (SEY) of samples irradiated in the accelerator. In the Project X upgrade the Main Injector will have its beam intensity increased by a factor of three compared to current operations. This may result in the beam being subject to instabilities from the electron cloud. Measured SEY values can be used to further constrain simulations and aid our extrapolation to Project X intensities. The SEY test-stand, developed in conjunction with Cornell and SLAC, is capable of measuring the SEY from samples using an incident electron beam when the samples are biased at different voltages. We present the design and manufacture of the test-stand and the results of initial laboratory tests on samples prior to installation.

## INTRODUCTION

A multi-MW proton facility has been established as a critical need for the High Energy Physics Programme of the USA by High Energy Physics Advisory Panel. Project X proposes to use the Fermilab Main Injector synchrotron (MI) as a high intensity proton source capable of delivering 2 MW beam power. The MI currently provides over 350 kW of beam power and is being upgraded, for the NOVA project, to provide 700 kW. Therefore, Project X will require a further factor three increase in intensity. Instabilities associated with beam loading and electron cloud effects are common issues for high intensity machines and these need to be studied if Project X is to successfully utilise the MI.

*Electron Cloud*

The electron cloud is an increase in the number of free electrons in an accelerator vacuum vessel. They can be generated in a number of ways, e.g., ionisation of residual gas particles in the vessel or emission from the vacuum vessel after impact from incident particles or synchrotron radiation. The electron cloud effect is generally an issue for machines with high currents or substantial synchrotron radiation, such as the LHC and the ILC damping rings, and possibly for Project X and the MI.

After being generated the cloud can also be amplified by the beam's electromagnetic field and can have various deleterious effects, including, increasing vacuum activity and detector backgrounds, depositing heat and generating beam instabilities. Simulations and modelling are required to predict the build-up and effect of electron clouds, therefore, it is necessary to compare these with measurements, especially before extrapolating to higher beam currents.

In the MI there have been a number of studies looking at the electron cloud, and related effects. For example, there have been attempts at measuring the electron cloud by studying the transmission of microwaves through the vacuum vessel [1]. As well as measuring the electron cloud quantifying and mitigating its build-up is also an important area of research. These studies are concerned with how the emission of secondary particles from the beam pipe is affected by the vessel material, coatings and the accumulated dose received from the beam.

*Secondary Electron Yield*

The SEY is simply the ratio of the number of secondary electrons emitted from a surface, $I_{SEY}$, to the number of electrons incident to that surface, $I_P$. Measuring the SEY of different samples and coatings has been done for many years, however the effect of the accumulated dose on the material in an accelerator environment has only recently been studied. Previous SEY studies have been done on samples exposed to a particle beam and then measured in an external lab [3]. This meant that the time between measurements was often several months, and the SEY as a function of accumulated dose was not well known. A series of in-situ measurements of the SEY have been underway for some time at Cornell [2]. These are with an electron beam and are also concerned with the effect of synchrotron radiation on different vessel materials. Similar measurements looking at the effect of the dose from protons are planned in the MI and described here.

## MEASUREMENT STAND

The measurement stand is similar to the one installed at Cornell, with some modifications to accommodate the MI vacuum vessel. There is also a port that will allow a residual gas analyser to be attached for measurements of the gas species present.

There are two measurements arms attached to the vacuum vessel, so two different samples can be exposed to the same dose for comparison. The samples are small curved circular pieces that sit flush with the beam pipe wall. When a measurement is to be taken they are retracted from the vacuum vessel wall into an electrically isolated arm. A remotely controlled Kimball Physics electron gun, also in the vacuum vessel, is directed towards the retracted samples. The gun fires a beam of electrons at the sample and the relevant currents are measured using a Keithley 6487 pico-ammeter to determine the SEY. During measurements the sample must be biased and this voltage is also applied by the Keithley pico-ammeter. Figure 1 and Figure 2 show the completed measurement stand.

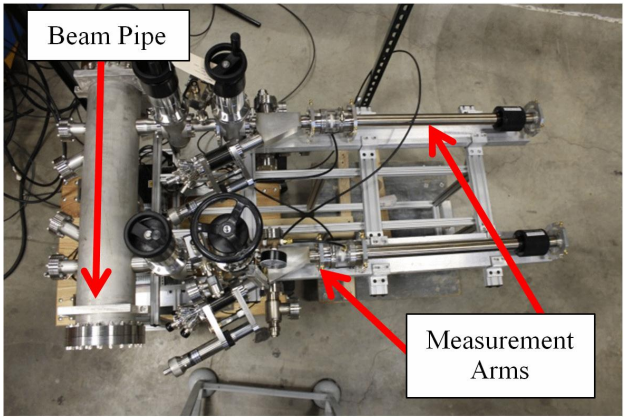

Figure 1: The SEY measurement stand.

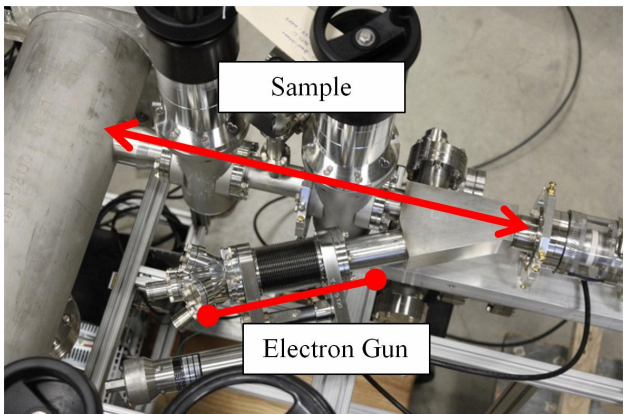

Figure 2: SEY measurement stand close-up. The sample, is retracted along the arm to be in line with the electron gun for measuring the SEY

## MEASUREMENTS

Before installation in the MI tunnel measurements of some aluminium samples were made in order to check the system. Some issues due to leakage currents were found resulting in some modifications to the design.

### Indirect Measurement of $I_{SEY}$

In this experimental set-up the vacuum vessel is at ground and so the SEY can only be measured indirectly. By biasing the sample at a high positive voltage any secondaries are recaptured, giving a measurement of $I_P$. A second measurement is taken with a small negative bias voltage. This repels the secondaries and also any electrons that are generated in other parts of the system [4] and gives a measurement of the total current, $I_T$. Care must be taken in getting the signs (positive or negative) of the properties measured in order to calculate the SEY, e.g. $I_{SEY}$ is opposite in sign to $I_P$. The SEY is calculated as:

$$SEY = \frac{I_T - I_P}{I_P} \quad (1)$$

### Controls and DAQ System

A schematic of the electrical system is shown in Figure 3. A PC is used to control the electron gun power supply, enabling the beam energy and current to be changed. The PC also controls the bias voltage applied to the sample and records the sample current during electron beam energy scans. Typically a scan automatically steps the electron gun energy from 20 eV to 1500 eV in increments of 10 eV. The process is controlled by a LabVIEW interface initially developed at Cornell [2] using existing software from Kimball Physics and Keithley. At each electron energy, the beam can be rastered across grid points to investigate different parts of the sample area. The focussing of the gun is set to give a constant beam size on the sample at each energy. To minimise error due to drift in the gun output current a second $I_P$ scan after the $I_T$ scan is performed and the two $I_P$ scans averaged.

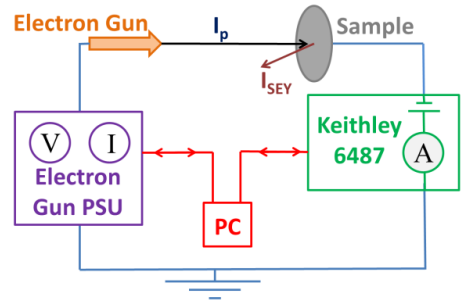

Figure 3: Schematic of experiment and DAQ, here a small negative voltage is applied to the sample, from the Keithley 6487, in order to repel secondary electrons

### Setting Bias Voltages

Figure 4 shows how the signal changes as the bias voltage goes from negative to positive. For the following measurements $I_T$ was measured with -20 V and $I_P$ with +150 V.

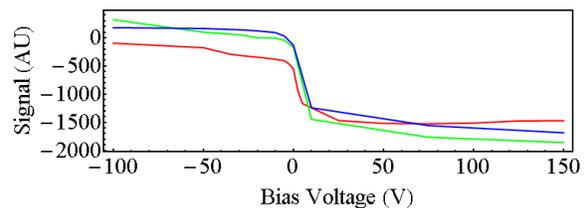

Figure 4: Ammeter signal changing the bias voltage for gun energies of 1500 (red) 500 (green) and 250 (blue) eV.

### Leakage Current

During measurements there are a number of places where current can leak from the sample to ground, generating a background signal that must be accounted for. These leakage currents principally occur if the ceramic, isolating the sample from the gun and beam pipe, becomes wet due to moisture in the atmosphere. Also, the screws connecting the support stands for the arm can leak a small current through them. During the initial tests measurements were taken with the ceramic exposed to the atmosphere, causing large leakage currents. For the next round of measurements the ceramic has been baked and is covered with a protective sleeve into which nitrogen is purging the atmosphere, Figure 5. The effect of the shroud on the leakage current is yet to be characterised. The screws for the support stands have been replaced with non-metallic ones which will also reduce the leakage current.

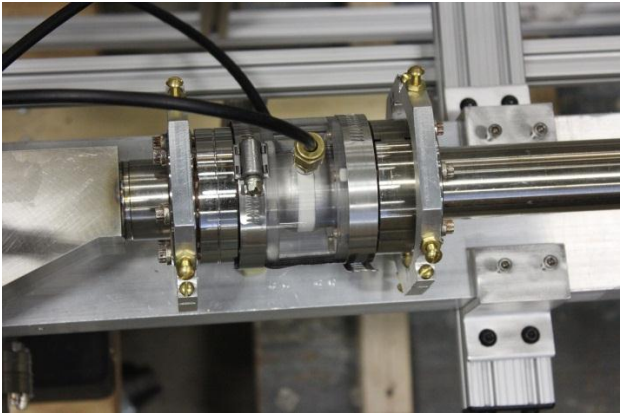

Figure 5: Close-up of ceramic and nitrogen shroud.

*Example Measurement*

The leakage currents at the two biasing voltages must be measured when there is no beam from the gun. For this set of measurements these typically had values of 55 pA and 388 pA for -20 V and +150 V respectively. These values are up to an order of magnitude greater than values obtained at Cornell when the gun ceramic is dry. Some example values for $I_P$, $I_T$ and then the SEY, calculated from Equation 1, are shown in Figures 6, 7 and 8.

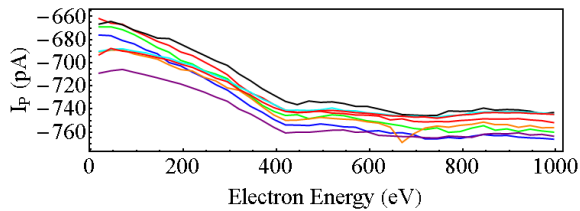

Figure 6: Example $I_P$ measurements without subtraction of leakage current.

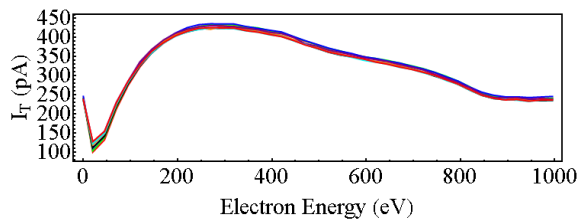

Figure 7: Example $I_T$ measurements without subtraction of leakage current.

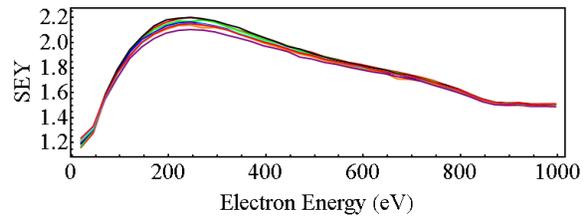

Figure 8: Example SEY.

## CONCLUSIONS AND FURTHER WORK

The electron cloud is an effect that must be understood for future high current operations of machines such as the Project X using the MI. One method of generating a cloud is through secondary emission of electrons from the vacuum vessel wall, therefore knowledge of the SEY of the vessel wall and how it varies with accumulated dose is necessary for modelling. A stand capable of in-situ measurements for the Fermilab MI has been designed, manufactured and in the process of being tested prior to installation.

The values for the SEY are a reasonable first attempt. For these first measurements the leakage current was high, and so a relatively high value of $I_P$ was also used to improve the signal to background ratio. However, this can lead to charging of the surface, and that could be the reason for the slight bumps seen in the spectrum at 600 to 800 eV. Improvements to the system to mitigate leakage currents effects have been incorporated. These include baking the ceramic and enclosing it in a shroud and nitrogen purge. These should enable $I_P$ to be reduced and any charging of the sample surface to be reduced.

There are also issues due to $I_P$ drifting during data acquisition. In theory it is possible to separately measure $I_P$ before each $I_T$, however when the bias voltage changes capacitances in the system can take of the order of minutes to damp during which no measurements can be taken. For this reason the time required to take a full scan, e.g. 25 eV to 1500 eV in 25 eV steps for a grid of 9 points on the sample would take ~4 hours, which is impractical. The possibility of enclosing the measurement arm in a Faraday cage is being considered. This should shield the experiment from many external fields and could allow the measurement time to be improved.

After testing the stand will be installed in the MI and allow for in-situ measurements of the SEY of different samples and different doses.